\documentclass[showpacs,preprintnumbers,amsmath,amssymb, eqsecnum,
twocolumn, tightenlines,
]{revtex4}

\usepackage{bm}

\usepackage{amsmath}

\usepackage{dcolumn}

   \usepackage[dvips]{graphicx}
\begin{document}

    \bibliographystyle{apsrev}
    
    \title {Reflection, radiation and interference for black holes}
    
    \author{M.Yu.Kuchiev}
    \email[Email:]{kuchiev@newt.phys.unsw.edu.au}

    \affiliation{ School of Physics, University of New South Wales,
      Sydney 2052, Australia}
    
%     \affiliation{
%      School of Physics, University of New South Wales,Sydney 2052,
%      Australia}

    \date{\today}

    \begin{abstract}
      Black holes are capable of reflection: there is a finite
      probability for any particle that approaches the event horizon
      to bounce back. The albedo of the black hole depends on its
      temperature and the energy of the incoming particle. The
      reflection shares its physical origins with the Hawking process
      of radiation, both of them arise as consequences of the mixing
      of the incoming and outgoing waves that takes place on the event
      horizon.
 \end{abstract}


    \maketitle

\section{Introduction} 
       \label{intro}
   
       Conventional intuitive arguments attribute two defining
       properties to black holes. The black holes are presumed to be
       unable to emit anything into the outside world, and are
       supposed to possess perfect absorption ability, i.e. be able to
       take in everything that comes close to their event horizon.
       There is, however, a known limitation for this simple intuitive
       picture that stems from thermodynamics that attributes
       temperature and entropy to black holes. The first indication
       that gravitational fields could have entropy came when
       investigation of Christodoulou \cite{christodoulou_1970} of the
       Penrose process \cite{Penrose_1969} for extracting energy from
       a Kerr black hole showed that there is a quantity which could
       not go down.  Hawking found \cite{hawking_1970} that it is
       proportional to the area of the horizon. Further research of
       Bardeen {\it et al} \cite{bardeen_carter_hawking_1970} demonstrated
       that black holes should obey laws similar to the laws of
       thermodynamics.  An important step made by Bekenstein
       \cite{bekenstein_1972,bekenstein_1973,bekenstein_1974} revealed
       that the area was actually the physical entropy. This
       suggestion was supported and enriched by the discovery of the
       Hawking radiation phenomenon \cite{Hawking_1974,Hawking_1975}.
       These works provided foundation for the thermodynamics approach
       to the black holes, for a recent review see Wald
       \cite{Wald_2001} and references therein, see also books of
       Frolov and Novikov \cite{frolov_novikov_1998}, Thorne
       \cite{thorne_1994}, and Chandrasekhar \cite{Chandrasekhar_1993}
       for comprehensive discussion of other black hole properties.
       
       The thermodynamics properties of black holes reveal that a
       black hole has the finite temperature $T$ and, correspondingly,
       is capable of radiation through the Hawking mechanism, in
       contradiction with the naive expectations.  In this work we
       address another property of black holes, their ability for an
       absorption.  From the first glance the Hawking mechanism of
       radiation supports an ability of black holes for perfect
       absorption. Indeed, the radiation spectrum of a black hole
       coincides with the spectrum of the perfect black body.  It is
       natural therefore to presume that the black hole is the black
       body, and as such should be a ideal absorber. However, we
       show that this perception is inadequate. The black holes are
       capable to reflect particles that come to their event horizon,
       and henceforth the black hole should not be considered as the
       perfect black body.  The fact that the radiation spectrum of
       the black hole coincides with the black body spectrum produces
       no contradiction. Interestingly, the radiation properties of the
       black hole and its reflection ability have one and the same
       physical origin.
       
       The classical description of the motion in the vicinity of the
       black hole horizon includes two types of trajectories.  There
       are the ingoing trajectories, they describe the motion towards
       the black hole center.  There are also the outgoing
       trajectories that lead out of the black hole center.
       Classically these two types of motion are quite different. If a
       particle, following the ingoing trajectory, approaches the
       event horizon, then it inevitably crosses it into the inside
       region.  After that it stays inside, there is no classically
       allowed way for it to switch to any outgoing trajectory that
       leads into the outside region, in full accord with intuitive
       feelings.  Discussing this point later we will use
       Fig. \ref{one} as an illustration to this statement.  The
       quantum description reveals a new rather unexpected feature of
       the problem. The event horizon produces a strong impact on the
       wave function of a probing particle.  The effect can be
       described in terms of interference or, equivalently, mixing of
       the incoming and outgoing waves. The incoming wave corresponds
       to the incoming classical trajectory; the correspondence
       between the waves and the trajectories works well in the
       vicinity of the event horizon because here the semiclassical
       description is applicable. The trajectory describes a smooth
       crossing of the horizon.  Similarly, the outgoing trajectory
       also describes the smooth transition through the
       horizon. However the quantum description reveals that some
       events that happen strictly on the horizon produce strong
       impact on the wave function, mixing the incoming wave with the
       outgoing one. In other words, the wave function of the incoming
       particle necessarily includes both the incoming and outgoing
       waves.
       
       The presence of the outgoing wave in the wave function has
       important physical implications.  One of them is the effect of
       reflection that is discussed in some detail in this work. The
       reflection means that there is a finite probability for an
       incoming particle to be reflected off the event horizon back,
       to the outside region. Another effect is the well known Hawking
       mechanism of radiation. It is demonstrated that the radiation
       can be considered as a consequence of the mentioned
       interference.  This new point of view provides an attractive
       physical picture that makes more clear some details on the
       radiation process

       To be more specific consider a particle in the outside region
       that approaches the black hole horizon. It is shown that there
       is a finite probability $\mathcal{P}$ for the particle to be
       reflected off the horizon,
   \begin{equation}
     \label{P}
{\mathcal P} = 
\exp\left( - \frac{\varepsilon-Q \Phi-J\omega}{kT} \right)~.
      \end{equation}
      This probability depends on the energy of the incoming particle
      $\varepsilon$, its charge $Q$, and its projection of the orbital
      momentum $J$ on the axis of rotation of the black hole. The
      essential parameters of the black hole that govern the process
      are the temperature $T$, the electric potential on the horizon
      $\Phi$, and the angular velocity of the horizon $\omega$.  The
      lower is the temperature, the stronger is the ability for
      reflection. 

      Notably, the probability of reflection (\ref{P}) coincides with
      the temperature factor that governs the Hawking radiation
      process, though the physical manifestation of the reflection
      differs from the radiation since the flux of the reflected
      particles is proportional to the magnitude of the incoming flux.
      Nevertheless, a similarity between the probability of reflection
      (\ref{P}) and the temperature factor is not accidental.  As was
      mentioned above, the reflection and radiation share the same
      physical origin, namely the interference of the incoming and
      outgoing waves.  A convenient way to prove the existence of this
      interference and to examine its magnitude is presented.
      Eq.(\ref{P}) is derived as a consequence of the interference for
      the reflection process, and a new derivation for the radiation
      process is discussed.

      Relativistic units $\hbar=c=1$ supplemented by the condition
      $2Gm = 1$ imposed on the gravitational constant $G$ and the black
      hole mass $m$ are used, if not stated otherwise. The
      Schwarzschild radius in these units reads simply $r_g =
      2Gm/c^2 \equiv 1$.

\section{Singularity of the  wave function on the horizon}
    \label{Singularity}

    Consider the static black hole described by the conventional
    Schwarzschild metric
    \begin{equation}
      \label{schw}
      ds^2 = -\left(1-\frac{1}{r}\right)dt^2 + \frac{dr^2}{1-1/r}
    +r^2 d\Omega^2~,
    \end{equation}
    where $d \Omega^2 = d \theta ^2 + \sin^2 \theta d\varphi^2$. The
    Hamilton-Jacobi classical equations of motion
    $g^{\kappa\lambda}\partial_\kappa S \partial_\lambda S = -\mu^2$
    for a particle with the mass $\mu$ in the metric (\ref{schw}) take
    the form
    \begin{equation}
      \label{schwS}
      \frac{ \dot{S}^2}{1-1/r} = \left(1-\frac{1}{r} \right)
      \left( \frac{\partial S}{\partial r} \right)^2
      +\frac{1}{r^2}
     \left( \frac{\partial S}{\partial \varphi} \right)^2 +\mu^2~.
       \end{equation}
       Separating the variables $S({\bf r},t)=-\varepsilon t+L
       \varphi +S(r)$ where $\varepsilon$ and $L$ are the energy and
       the momentum of the particle, $\varphi$ is its azimuthal
       angle, one finds the radial action
       \begin{equation}
         \label{schwS(r)}
         S(r) =\mp \int^r \left[ \varepsilon^2 -
        \left(\mu^2+\frac{L^2}{r} \right) 
\left(1-\frac{1}{r}\right) 
         \right]^{1/2} \frac{dr}{1-1/r}~.
       \end{equation}
       In the vicinity of the black hole horizon $r \rightarrow 1$,
       which plays an important role in the following discussion, the
       action (\ref{schwS(r)}) simplifies
       \begin{equation}
         \label{schwLn}
         S(r) = \mp \, \varepsilon \ln(r-1)~,
       \end{equation}
       that gives $S({\bf r},t) = - \varepsilon t \pm \varepsilon \ln
       (r-1) +L\varphi$. The corresponding equation of motion $
       \partial_\varepsilon S( {\bf r},t)=0 $ yields the radial
       trajectories
       \begin{equation}
         \label{r-1}
         r =1 + \exp(\mp \,t)~.
       \end{equation}
       The signs minus and plus in
       Eqs.(\ref{schwS(r)}),(\ref{schwLn}),(\ref{r-1}) correspond to
       the incoming and outgoing trajectories respectively. These
       equations are conveniently written for the outside region $r>1$
       (the inside region is discussed in Section \ref{interference}).
       It is important that the classical action for a probing has the
       logarithmic singularity (\ref{schwLn}) on the horizon.  The
       coefficient in front of the logarithm function is equal to the
       energy of the particle $\varepsilon$ ($\varepsilon r_g /c$ in
       absolute units) that plays an important role in what follows,
       eventually finding its way into the exponential function in
       (\ref{P}).  Importantly, the logarithmic singularity is an
       invariant property of the action, it persists even in those
       coordinates that eliminate the singularity of the metric on the
       horizon.  For example, in Kruskal coordinates U,V
       \cite{kruskal_1960} (for a comprehensive discussion of the
       Kruskal coordinates see Ref.\cite{misner_thorne_wheeler_1973})
       \begin{eqnarray}
         \label{U}
        U = &-& \sqrt{r-1}  \,\,\exp[\,(r-t)/2\,]~, 
         \\ \label{V}
        V = & \,\,\,\,&
\sqrt{r-1}  \,\,\exp[\,(r+t)/2\,]~, 
       \end{eqnarray}
       the metric $ds^2 = - 4 dUdV \exp(-r)/r +r^2 d\Omega^2$ is
       regular on the horizon, which is described by conditions $U=0$
       or $V=0$, but the logarithmic singularity
       of the action remains intact, it can be conveniently presented
       as
       \begin{eqnarray}
\label{lnV}
         S({\bf r},t) \simeq &  -&\varepsilon \ln (V^2)~,\quad \quad 
{\mathrm {for}}\quad
         U\rightarrow 0~,
\\          \label{lnU}
         S({\bf r},t) \simeq &\!\!\!& \varepsilon \ln (U^2)~,\quad \quad 
{\mathrm {for}}\quad
         V\rightarrow 0~.
      \end{eqnarray}
      The classical action allows one to find the semiclassical wave
      function $\Phi({\bf r},t)$ that describes the coordinate motion
      of the particle (leaving aside possible spin variables).
      Separating the variables, $\Phi({\bf r},t)= \exp(-i\varepsilon
      t) Y_{LM}(\theta,\varphi)\phi(r)$, where
      $Y_{LM}(\theta,\varphi)$ is the conventional spherical function
      describing the motion with the orbital momentum $L$ and its
      projection $M$, one presents the semiclassical radial wave
      function $\phi(r)$ as
      \begin{eqnarray}
        \label{phi}
        \phi(r) \propto \exp[\,i S(r)\,] \simeq \exp[\,\mp i\,\varepsilon 
        \ln(r-1)\,] ~.
      \end{eqnarray}
      It is verified below, see after (\ref{phi''}), that the
      preexponential factor in (\ref{phi}) is a constant, which we
      chose to be unity.  Thus the singularity of the action at $r=1$
      results in the corresponding singularity of the wave function.
      
      In order to scrutinize this result one needs to assess validity
      of the semiclassical description in the vicinity of the horizon.
      To this end consider the wave function $\Phi({\bf r},t)$ as a
      solution of the Klein-Gordon equation for the scalar field.
      From (\ref{schw}) one finds that the radial wave function
      $\phi(r)$ satisfies the equation
      \begin{eqnarray}
        \label{phi''}
&&\phi'' + \left(\frac{1}{r}+\frac{1}{r-1} \right) \phi'\\ \nonumber && +
\frac{1}{1-1/r}\left( \frac{\varepsilon ^2}{1-1/r} -\mu^2 
- \frac{L(L+1)} {r^2} \right)  \phi = 0~.
      \end{eqnarray}
      In the vicinity of the horizon $r=1$ the solution can be
      approximated by $\phi(r)\simeq (r-1)^\eta$ where (\ref{phi''})
      yields $\eta=\pm i\varepsilon$. The agreement with the
      semiclassical result (\ref{phi}) supports its validity and
      verifies that the preexponential factor in (\ref{phi}) is,
      indeed, a constant.  It is instructive also to look at the
      singularity of the wave function (\ref{phi}) from the point of
      view of the conventional Schr\"odinger-type equation. Making the
      substitution $\phi(r) \rightarrow \psi(r)= [\,r(r-1)\,]^{1/2}
      \phi(r)$ one rewrites (\ref{phi''})
      \begin{eqnarray}
        \label{p2}
        p^2 \psi(r) = - \psi''(r)+U(r)\psi(r)~,
      \end{eqnarray}
where
\begin{eqnarray}
  \label{UU}
U(r) = &-&\frac{1}{ (r-1)^2}\left(\varepsilon^2 + \frac{1}{4r^2}\right)
\\ \nonumber
&-& \frac{1}{r-1}\left( \varepsilon^2 + p^2-\frac{L(L+1)}{r} \right). 
    \end{eqnarray}
    Eq.(\ref{UU}) has the form of the Schr\"odinger-type equation if we
    consider $U(r)$ as an effective, energy-dependent potential and
    accept the momentum $p^2$ on the left-hand side as the eigenvalue.
    For $r\rightarrow 1$ the potential exhibits a notable feature
    \begin{eqnarray}
      \label{r2}
      U(r)\rightarrow -\frac{\varepsilon^2+1/4}{(r-1)^2}~.
    \end{eqnarray}
    It is well known in nonrelativistic quantum mechanics
    \cite{landau_lifshits_1977} that in the potential $U(z) =
    -U_0/z^2$ for $U_0 > 1/4$ the wave function collapses to the point
    $z=0$. Since the necessary inequality is obviously satisfied in
    (\ref{r2}), $\varepsilon^2 + 1/4>1/4$, one conclude that (\ref{p2})
    indicates the collapse of the wave function on the event horizon
    $r=1$.  This fact could be interpreted as the absorption of the
    particle by the black hole.  Thus from the first sight the quantum
    description seems to agree with classical arguments based on the
    incoming trajectory in (\ref{r-1}) that converges to the event
    horizon, supporting also the intuitive perception of the black
    hole as an ideal absorber.  However, more careful discussion
    below exposes limitations of this point of view.
      
    Summarizing, it is demonstrated that the wave function $\phi(r)$ has
    a singularity (\ref{phi}) on the event horizon.

\section{Reflection}
      \label{reflection}
      
      Consider a particle that approaches the event horizon of the
      black hole. Let us describe its radial motion with the help of
      the wave function $\phi(r)$. According to (\ref{phi}) the wave
      function in the vicinity of the horizon can be written as
\begin{eqnarray}
  \label{R}
  \phi(r) = \exp[ - i\,\varepsilon \ln(r-1) ] +
           {\mathcal R} \exp[ i\,\varepsilon \ln(r-1)]~.
    \end{eqnarray}
    The first term here describes the proper incoming wave, while the
    second one, that presents the outgoing wave, is written in order
    to allow for an opportunity of the possible interference of the
    incoming and outgoing waves in the wave function. If this
    interference takes place, i. e. if ${\mathcal R} \ne 0$, then the
    outgoing wave in (\ref{R}) clearly indicates that there is
    the probability for the incoming particle to be reflected on the
    horizon. The unitarity condition implies $| {\mathcal R}| \le 1$.
    Moreover, intuitively one would expect the reflection coefficient
    in (\ref{R}) to be zero, ${\mathcal R}=0$.  Such assumption would
    agree with a naive perception of the black hole as a perfect
    absorber. However, in order to verify, approve or reject this
    intuitive claim (we will reject it, in fact) one needs to examine
    carefully what happens with the wave function on the horizon.

    Straightforward discussion of events that happen strictly at $r=1$
    faces an obstacle produced by the singular nature of the wave
    function (\ref{R}) at this point.  Fortunately, one can avoid
    discussion of the events that take place strictly on the horizon
    r=1 using the analytical continuation of the wave function in the
    vicinity of this point.  Consider the distance from the horizon $z
    = r-1$, treating $z$ as a complex variable. The wave function
    (\ref{R}) is explicitly analytical in $z$, except for the power-type
    singularity at $z=0$ that induces a cut emerging from this point
    on the complex plane $z$. Let us take $r$ in the outside region of
    the black hole in a close vicinity of the event horizon, which
    means that $0<z\ll 1$ , and examine what happens with the wave
    function when one rotates $z$ in the complex $z$-plane over an
    angle $2\pi$ clockwise (the anti-clockwise rotation is forbidden,
    see discussion after (\ref{T})).  We can keep $|z|$ small, $|z|
    \ll 1$, during this rotation, thus justifying validity of the
    semiclassical wave function (\ref{R}).  This analytical
    continuation necessarily incorporates a crossing of the cut on the
    complex plane. Therefore after finishing this rotation and
    returning to a real, physical value $z >0$, the wave function
    acquires a new value on its Riemann surface, let us call it $\phi^{
      (2\pi) }(r)$. A procedure of this type is usually referred to as a
    monodromy. In our case the monodromy can be read of (\ref{R})
    \begin{eqnarray}
      \label{2pi}
\!\! \!
\phi^{(2\pi)}(r) \!=\! {\varrho} \exp [-i\varepsilon \ln (r\!-\!1) ]
\!+ \! \frac{\mathcal R}{\varrho} \exp [i\varepsilon \ln (r\!-\!1) ],
   \end{eqnarray}
   where $\varrho = \exp(-2\pi \varepsilon)$.  The analytically
   continued function $\phi^{(2\pi)}(r)$ satisfies the same real
   differential equation as the initial function $\phi(r)$. Moreover,
   one has to expect that the wave function satisfies the same
   normalization conditions as the initial wave function $\phi(r)$.
   This needs that one of the coefficients in (\ref{2pi}), either
   $\varrho$, or $\mathcal{R}/\varrho$ should have an absolute value
   equal to unity. Since $\varrho <1$, we deduce that
   $| \mathcal{R} |/\varrho =1$ thus concluding that
   \begin{eqnarray}
     \label{R=}
     |\mathcal{R}| = \exp \left( - 
\frac{2 \pi r_g \varepsilon}  {\hbar c} \right)~,
     \end{eqnarray}
     where the conventional units are used to make the result more
     transparent. We see that the reflection coefficient is nonzero.
     In other words, the black hole is capable of reflection, in a
     notable contradiction with the naive intuitive expectations.
     There exists a way to scrutinize (\ref{R=}). Recall again that
     $\phi^{(2\pi)}(r)$ satisfies the same differential equation as
     the wave functions $\phi(r)$ and $\phi^*(r)$.  One should be able
     therefore to present $\phi^{(2\pi)}(r)$ as their linear
     combination $\phi^{(2\pi)}(r) = \alpha \phi(r) + \beta
     \phi^*(r)$. From (\ref{R}) and (\ref{2pi}) we derive that this
     linear combination really exists, having a simple form
     \begin{eqnarray}
       \label{beta}
       \phi^{(2\pi)}(r) = \beta \phi^*(r)~,\quad |\beta|=1~. 
     \end{eqnarray}
     Thus the physical picture presented is self-consistent.
     Additional support for the results discussed provide
     Refs.\cite{kuchiev_03,kuchiev2_03} that suggest alternative
     approaches leading to Eq.(\ref{R}).  Ref.\cite{kuchiev_03} argues
     that Eq.(\ref{beta}) expresses the fundamental symmetry of the
     space-time. Starting from this symmetry condition,
     Ref.\cite{kuchiev_03} derives Eq.(\ref{R=}).
     Ref.\cite{kuchiev2_03} adopts another point of view. It relies
     more heavily on the dynamics of the system represented by the
     wave equation (\ref{phi''}). This work claims that an accurate
     treatment of the solution of this equation in the vicinity of its
     three singular points, $r=0,~r=1,~r=\infty$, leads to
     Eq.(\ref{R=}). 
     
     From (\ref{R=}) we see that the incoming and the outgoing waves
     interfere in the wave function (\ref{R}). Correspondingly, there
     is the reflection.  The probability of reflection can be found as
     $\mathcal{P} = |\mathcal{R}|^2$, that in view of (\ref{R=}) gives
     $\mathcal{P}=\exp(- 4 \pi \varepsilon )$ in full accord with
     (\ref{P}) proclaimed in Section \ref{intro}.  The parameter $T$
     that appears in (\ref{P}) arises from the coefficient in front of
     the logarithmic function in (\ref{schwLn})
     \begin{eqnarray}
       \label{T}
       kT = \frac{\hbar c}{4 \pi r_g}~,
     \end{eqnarray}
     (absolute units). Notably, it proves be equal to the Hawking
     temperature of the black hole. Applying (\ref{P}) one should
     remember, of course, that the electric potential and rotational
     frequency for the Schwarzschild case are absent, $\Phi=\omega=0$.
     
     Let us return back to examine why it was necessary to use
     specifically the clockwise rotation when the analytical
     continuation of the wave function (\ref{P}) in the complex
     $z$-plane was fulfilled. A simplified answer to this question is
     that an attempt to use the counter-clockwise rotation leads to a
     self-contradiction. Trying it, i.e. making the anti-clockwise
     rotation, one arrives to the result similar to (\ref{2pi}), but
     with the different coefficient $\varrho'$ instead of $\varrho$,
     $\varrho \rightarrow \varrho'=1/\varrho = \exp(2\pi\varepsilon)$.
     Proceeding further one would be forced to conclude that the
     reflection coefficient is $|\mathcal{R}'| = \exp(2\pi
     \varepsilon)$, which comes into an obvious contradiction with the
     unitarity condition for the reflection that specifies
     $|\mathcal{R}'|\le 1$. It is a known, common feature of the
     semiclassical wave function that different ways for its
     analytical continuation lead to different results, and one needs
     to choose carefully an appropriate way of continuation (\ref{R}).
     To outline deeper roots of this problem in our specific case it
     is convenient to use Kruskal coordinates (\ref{U}),(\ref{V}). It
     is known from the analysis of Hartle and Hawking
     \cite{hartle_hawking_1976} that the propagator of the scalar
     particle in the Schwarzschild metric is an analytical function of
     $U$ and $V$ in the upper half-plane of the complex $U$-plane and
     in the lower half-plane of the complex $V$-plane.  In terms of
     the variable $z$ this means that the propagator remains an
     analytical function when it is continued from the real semi-axes
     $z>0$ in the clockwise direction over the angle $2\pi$.  There is
     a slight distinction in our case. Our analysis relies on the wave
     function, while the work \cite{hartle_hawking_1976} refers to the
     properties of the propagator. However, the analytical properties
     of the wave function are similar to those of the propagator. We
     conclude that the analytical continuation used in derivation of
     (\ref{2pi}) is justified. In contrast, an attempt to use the
     analytical continuation rotating $z$ from the region $z>0$ in the
     counter-clockwise direction should meet a difficulty.  It really
     does, as demonstrated an attempt discussed above.
       
     Let us summarize the main ideas used in derivation of (\ref{R})
     and (\ref{R=}).  We verified firstly that in the vicinity of the
     event horizon the semiclassical approximation is valid, while the
     classical action possesses the logarithmic singularity
     (\ref{schwLn}). Then we analyzed consequences of this singularity
     for the wave function using its analytical continuation in the
     complex $r$-plane in the vicinity of the event horizon.
       
     The mixing coefficient (\ref{R=}) has interesting properties.
     Firstly, it vanishes in the classical limit $\hbar \rightarrow
     0$.  Therefore it has no analog in the classical description, in
     this sense it is an {\em unexpected} result, making its physical
     consequences also unexpected.  Moreover, even in the quantum
     picture one needs to make an effort to distill it.  Remember that
     from the first sight the gravitational field is taken into
     account in the radial incoming and outgoing waves in (\ref{phi})
     adequately and completely. However, the result obtained in
     (\ref{R=}) indicates that there exists some part of the
     gravitational interaction that remains unaccounted for in these
     wave functions, call it an ``additional'' interaction. Its
     existence can be suspected since the wave functions in
     (\ref{phi}) are singular on the horizon.  The argument in favor
     of a possible additional interaction can also be drawn from the
     singular nature of the potential (\ref{r2}) in the
     Schr\"odinger-type equation (\ref{p2}).  Dealing with the
     potential that is so singular one can suspect that some part of
     this potential, the $\delta$-term localized on the horizon, i. e.
     proportional to $\delta(r-1)$ or its derivatives, may remain
     unaccounted for.  This part, if exists, would mix the incoming
     and outgoing waves.  The suspected additional singularity of the
     potential has to be moderated by the causality conditions. This
     means that one can consider the adiabatic "switched off" of the
     black hole in the distant past. A precise mechanism for this
     switching off is not relevant, it is sufficient to assume only
     that this procedure is foreseeable.  When the black hole is
     switching off, the singularity on the horizon disappears. That is
     why one can use a method of the analytical continuation. The
     analytical conditions are closely related to the causality
     principle, incorporating its consequences.  Developing this
     argument we take the wave function (\ref{R}) close to the
     horizon, but outside it, $|r-1| \ll 1,~ r>1$. In this region the
     gravitational field that exists for $r\ne 1$ is included in each
     of the two waves on the right-hand side almost completely, except
     for the possible $\delta$-term that can manifest itself only
     through the mixing of these two waves.  Imposing the causality
     condition through the analytical continuation we prove that this
     mixing really takes place.  In our derivation the distance from
     the horizon can be made arbitrary small, $|r-1|\rightarrow 0$.
     This indicates that the mixing of the incoming and outgoing waves
     in (\ref{R}) originates from those events that are localized on
     the horizon $r=1$, in accord with the expectation that the effect
     is due to a $\delta$-term that is missed in the potential
     (\ref{UU}).
       
     The logarithmic singularity of the radial action plays a central
     role in the above derivation. It is instructive to compare this
     singularity with the behavior of the classical trajectory.
     Recalling the classical equations of motion $\partial_\varepsilon
     S({\bf r},r)=0$ one observes that the well-known exponential
     function in the trajectory (\ref{r-1}) and the logarithmic
     singularity of the action are simply one and the same property
     expressed by two different means. Thus the reflection property
     found is closely related to the exponential-type behavior of the
     classical trajectory.
       
       The result given in (\ref{R}),(\ref{R=}) indicates that the
       horizon is capable of reflecting the incoming particle with
       probability given in (\ref{P}). By the same token it means that
       the probability for the incoming particle to cross the horizon
       $\mathcal{P}_\mathrm{cr}$ penetrating into the inside region is
       less than unity
       \begin{eqnarray}
         \label{cro}
         \mathcal{P}_\mathrm{cr}=1- \mathcal{P}~.
       \end{eqnarray}
       We will use (\ref{cro}) in Section \ref{interference}
       discussing the Hawking radiation process.  Summarizing, it is
       demonstrated that the black hole is capable to reflect
       particles that come to its horizon, the reflection probability
       satisfies (\ref{P}).

      \section{Reflection by different  types of  black holes}
      \label{reflectionGen}
      
      This Section extends the results derived above for the
      Schwarzschild black hole to other, more complex types of black
      holes. We rely on the step-by-step approach considering first
      the Reissner-Nordstr\"om solution, then the Kerr solution and
      only after that the general Kerr-Newman solution. The reader
      familiar with these solutions may prefer to go directly to
      Section \ref{Kerr-Newman} that discusses the general case.

      \subsection{Charged  black holes}
      Consider the Reissner-Nordstr\"om black hole with the mass $m$
      and charge $q$. Its metric is given by
   \begin{equation}
      \label{RN}
      ds^2 = -\left(1-\frac{1}{r} +\frac{q^2}{r^2} 
\right)dt^2 + 
      \frac{dr^2}{1-1/r + q^2/r^2}+r^2 d \Omega^2~.
    \end{equation}
    The Hamilton-Jacobi equation for the particle with the mass $\mu$,
    charge $Q$ and orbital momentum $L$ for the metric (\ref{RN})
    reads
    \begin{eqnarray}
      \label{RNHJ}
      \frac{ \left( \dot{S} - Q \Phi \right) ^2 }
      {1-1/r + q^2/r^2 } =
      \left(1-\frac{1}{r} +\frac{q ^2 }{r^2} \right)
      \left( \frac{\partial S}{\partial r} \right)^2
\\ \nonumber
      +\frac{1}{r^2}
     \left( \frac{\partial S}{\partial \varphi} \right)^2 +\mu^2~,
    \end{eqnarray}
    where $\Phi(r)= q/r$ is the black hole electric potential (compare
    Eq.(\ref{schwS}).  Separating the variables $S({\bf r},t) =
    -\varepsilon t + L \varepsilon + S(r)$ one derives
    \begin{eqnarray}
\nonumber     
             \!\!  S(r) \! &=& \! \mp \!  \int \!\! 
\left[ (\varepsilon \!- \! Q \Phi(r))^2 \!-\!
       \left(\mu^2\!+\!\frac{L^2}{r} \right) 
      \left(1\!-\! \frac{1}{r} \!+\! \frac{q^2 }{r^2}\right) 
         \right]^{1/2} 
\\ \label{RNS(r)}
&& \times \frac{dr^2}{1-1/r + q^2/r^2 }~.
    \end{eqnarray}
    The poles of $g_{rr} = 1-1/r+q^2/r^2$ are located on
    two spherical surfaces with radiuses
    \begin{eqnarray}
      \label{sqrt}
    r_{\pm} = \frac{1}{2}\pm \sqrt{ \frac{1}{4}-q^2 }~.
    \end{eqnarray}
    The largest of them with the radius $r_+$ represents the black hole
    horizon. In the vicinity of the horizon $r\rightarrow r_+$ one
    finds from (\ref{RNS(r)})
    \begin{eqnarray}
      \label{zetaln}
      S(r) \simeq \mp\,\zeta \, \ln (r-r_+)~,
    \end{eqnarray}
    where
    \begin{eqnarray}
      \label{zeta}
\zeta = [ \, \varepsilon -Q \Phi(r_+) \, ] 
\frac{r_{+} ^2 }{ r_{+}  - r_{-} }~.
    \end{eqnarray}
    In analogy with (\ref{schwLn}) the action (\ref{zetaln}) possesses
    the logarithmic singularity. We can therefore follow the way paved
    by Eqs.(\ref{R}),(\ref{2pi}) and (\ref{R=}). Firstly we construct the
    wave function
      \begin{eqnarray}
      \label{phidzeta}
      \phi(r) = \exp[ - i\,\zeta \ln(r-1) \, ] +
           {\mathcal R} \exp[ \,i\,\zeta \ln(r-1) \, ],
      \end{eqnarray}
      that describes the radial motion of the particle in the vicinity
      of the event horizon. Then, introducing the variable $z = r -
      r_{+}$, and assuming that $z>0$, $ |z| \ll r_{+} - r_{-} $, i.
      e. taking $r$ in the external region in a close vicinity of the
      event horizon, we make the analytical continuation rotating $z$
      in the complex plane $z \rightarrow \exp(- i \gamma)z~,\gamma
      \ge 0$, eventually taking $ \gamma = 2\pi $. This procedure
      gives the coefficient of reflection $ \mathcal{R} = \exp(-2 \pi
      \zeta)$ and the probability of reflection $ \mathcal{P} =|
      \mathcal{R} |^2 = \exp (-4 \pi \zeta )$.  The latter result
      agrees with (\ref{P}), where the value for the parameter $T$
      follows from (\ref{zetaln}),(\ref{zeta})
      \begin{eqnarray}
        \label{RNT}
        kT = \frac{\hbar c}{4 \pi} \, \frac{ r_{+} -r_{-} }{r_{+}^2}~,
      \end{eqnarray}
      (absolute units). It proves equal to the Hawking temperature
      of the charged black hole.

      \subsection{Rotating black holes}
      
      Consider the Kerr black hole that possesses the mass $m$ and the
      spin $j$ that is conveniently parameterized by $a=j/m$. The
      Kerr metric in the Boyer-Lindquist coordinates reads
      \begin{eqnarray}
        \label{Kerr}
%ds^2 =&-& \frac{\Delta}{\rho}(\, dt-a\sin^2 \theta \,d\varphi\,)^2 \\
%\nonumber
%&+& \frac{\sin^2 \theta}{\rho^2}\,
%[ (r^2+a^2)\, d\varphi -a \,dt\,]^2 + 
%\frac{\rho^2}{\Delta }\, dr^2 + \rho^2 \, d \theta^2\,.
ds^2 =
&-&\frac{\Delta}{\rho^2} (\,dt-a \sin^2 \theta \,d\varphi\,)^2 \\
\nonumber
&+& \frac{\sin^2 \theta}{\rho^2} 
\Big[ \,(r^2+a^2)\,d\phi -a \,dt\,\Big]^2 \\
\nonumber
&+&\frac{\rho^2}{\Delta }\, dr^2 + \rho^2 \, d \theta^2\,.
      \end{eqnarray}
      Here $\Delta =r^2-r+a^2$ and $\rho^2 = r^2 + a^2 \cos^2 \theta$.
      The nodes of $\Delta $, are located on the two spheres with
      radiuses
      \begin{eqnarray}
        \label{rpm}
 r_{\pm} = \frac{1}{2}\pm \sqrt{ \frac{1}{4}-a^2 }~,
      \end{eqnarray}
      the largest of which represents the black hole horizon. The
      Hamilton-Jacobi equations of motion for the metric (\ref{Kerr}),
      \begin{eqnarray}
%\nonumber
%&& \frac{1}{\Delta} 
%\left( r^2 + a^2+ 
%\frac{ra^2}{\rho^2} \sin^2 \theta  \right)^2 \dot{S}^2 
%-\frac{\Delta}{\rho^2} 
%\left( \frac{ \partial S} {\partial r} \right)^2
%\\ \nonumber
%&-&\frac{1} {\rho^2} 
%\left(  \frac{\partial S}{\partial \theta} \right)^2
%-\frac{1} {\Delta \sin^2 \theta} 
%\left(  1-\frac{r}{\rho^2} \right) 
%\left(  \frac{\partial S} {\partial \varphi} \right)^2 
%\\        \label{KerrHJ}
%&+& 2 \left( \frac{ra}{\rho^2 \Delta} \right) 
%\frac{ \partial S}{\partial \varphi} \, \dot{S} = \mu^2~,
\nonumber
&& \frac{1}{\Delta \,\rho^2} \left[\,
( r^2 + a^2)\,
\frac{\partial S}{\partial t} +a \,\frac{\partial S}{\partial \varphi}
\,\right]^2  \\ \nonumber
&-& \frac{1}{\rho^2 \sin^2 \theta}\left[ 
a\,\sin^2 \theta \,\frac{\partial S}{\partial t} 
- \frac{\partial S}{\partial \varphi} \right]^2 \\ \label{KerrHJ}
&-&\frac{\Delta}{\rho^2}\left(\frac{\partial S}{\partial r}\right)^2
-\frac{1}{\rho^2}\left(\frac{\partial S}{\partial \theta}\right)^2
=\mu^2~,
      \end{eqnarray}
      allow the full separation of variables $S({\bf r},t) =
      -\varepsilon t + J \varphi + \Sigma(\theta) + S(r)$ that 
produces the following result for the radial action $S(r)$
\begin{eqnarray}
  \label{KerrS(r)}
%\!\! \Delta \left( \frac{dS(r)}{dr}\right)^2 \! \! \!  -\!
%\frac{1}{\Delta} [ (r^2 \! + \! a^2) \varepsilon  - Ja ]^2 
%\! +  \! \mu^2 r^2 \! =-\! K.
&&S(r) = \int\Delta^{-1} \sqrt{R} \,\,dt~, \\ \label{RR}
&&R    = P^2 - \Delta\,[\,\mu^2 r^2+K\,]~, \\ \label{PP}
&&P    = \varepsilon\,(r^2+a^2)-a J~,
      \end{eqnarray}
      Here $J$ is the conserved projection of the orbital momentum of
      the particle on the axis of rotation of the black hole, and $K$ is
      an additional ("accidental") integral of motion. In the vicinity
      of the horizon $r\rightarrow r_{+}$, $\Delta \rightarrow 0$, one
      finds from (\ref{KerrS(r)}) that $S(r)$ has a logarithmic
      singularity that satisfies (\ref{zetaln}) in which the parameter
      $\zeta$ equals
      \begin{eqnarray}
        \label{KerrZeta}
        \zeta = (\varepsilon -J \omega) \frac{ r_{+}^2 + a^2}
        { r_{+} - r_{-} }~.
      \end{eqnarray}
      Here $\omega = a/( r_{+}^2+a^2 )$ is the frequency of rotation
      of the black hole horizon. Using the method well-discussed above
      we derive from the logarithmic singularity that the rotating
      black hole is capable of reflection, the probability of
      reflection is given by (1), in which (\ref{KerrZeta}) predicts
      for the parameter $T$
      \begin{eqnarray}
        \label{KerrT}
        kT =\frac{\hbar c}{4 \pi} \, \frac{r_{+}-r_{-} }
        {r_{+}^2 + a^2 }~,
      \end{eqnarray}
      (absolute units) that coincides with the temperature of the
      rotating black hole. 
      
\subsection{Charged-rotating black holes}
      \label{Kerr-Newman}
      Consider the general case of the Kerr-Neumann black hole that
      possesses both the charge $q$ and the spin $j$. The Kerr-Newman
      metric in the Boyer-Lindquist coordinates is described by
      Eq.(\ref{Kerr}) in which the parameter $\Delta$ reads
\begin{eqnarray}
  \label{Delta}
\Delta = r^2-r+a^2+q^2~.
      \end{eqnarray}
      The nodes of $\Delta$ are located on the spheres with radiuses
      \begin{eqnarray}
        \label{rpmKN}
 r_{\pm} = \frac{1}{2}\pm \sqrt{ \frac{1}{4}-a^2 -q^2}~,
      \end{eqnarray}
      the largest of which represents the black hole horizon.  The
      electromagnetic field of the black hole is described by the
      vector potential $A_\mu dx^\mu =
      -(qr/\rho^2)(dt-a\sin^2\theta\,d\varphi))$. The Hamilton-Jakobi
      equations of motion for a charged particle moving in the
      gravitational and electromagnetic fields created by a black
      hole allow the full separation of variables, see e.g. p.901 of
      Ref.\cite{misner_thorne_wheeler_1973}. One promptly finds that
      the radial action is described by
      Eqs.(\ref{KerrS(r)}),(\ref{RR}) in which the parameter $P$
      equals
      \begin{eqnarray}
        \label{PPKN}
      P  = \varepsilon\,(r^2+a^2)-a J -qQr~.
      \end{eqnarray}
      From Eqs.(\ref{KerrS(r)}),(\ref{RR}),(\ref{PPKN}) we find that on
      the horizon $r\rightarrow r_+$ the action has the logarithmic
      singularity 
    \begin{eqnarray}
      \label{zetalnKN}
      S(r) \simeq \mp\,\zeta \, \ln (r-r_+)~,
     \end{eqnarray}
     where
    \begin{eqnarray}
      \label{zetaKN}
\zeta = [ \, \varepsilon -Q \Phi(r_+) -J \omega \, ] 
\,\frac{r_{+} ^2+a^2 }{ r_{+}  - r_{-} }~.
     \end{eqnarray}
     Here $\Phi(r_+) = qQr_+/(r_+^2+a^2)$ is the potential describing
     interaction of the particle with the electromagnetic field of the
     black hole on the horizon.  Using Eq.(\ref{zetaKN}) and applying
     the method well described above one proves that the reflection
     probability for the Kerr-Newman black hole is given by (\ref{P}).
     The parameter $T$ that appears in (\ref{P}) satisfies
     Eq.(\ref{KerrT}) with $r_\pm$ from Eq.(\ref{rpmKN}); this $T$
     coincides with the temperature of the Kerr-Newman black hole.
     Setting in Eqs.(\ref{KerrT}),(\ref{rpmKN}) either $q$, or $j$, or
     both of them to zero one returns to the case of the Kerr black
     hole, the Reissner-Nordstr\"om black hole, and the Schwarzschild
     black hole respectively.
      
     We relied above on the semiclassical approach.  Eq.(\ref{rpmKN})
     can be improved to account more accurately for the quantum
     properties of the momentum $j$ by substituting $j^2\rightarrow
     j(j+1)$ in $a^2$ in (\ref{rpmKN}). This issue becomes important
     when quantum properties of the black hole itself are considered,
     see recent works of Bekenstein devoted to this subject
     \cite{bekenstein_2002_a,bekenstein_2002_b}. However, for the
     purpose of this work this subtlety is not essential.

     We discussed in this Section several types of black holes that
     possess either the charge, or momentum or both, verifying that in
     each and every case the black hole is capable of reflection.  Our
     most general result, which is presented for the Kerr-Newman
     solution, is described in
     Eqs.(\ref{P}),(\ref{KerrT}),(\ref{rpmKN}). There are known a
     number of more sophisticated solutions for the black holes with
     hair, see the review \cite{volkov_gal'tsov_1999}, but we leave
     them outside the scope of the present work.
     
\section{Interference,  reflection and radiation}
      \label{interference}
      
      Let us show that the reflection ability of the black hole and
      the phenomenon of Hawking radiation have the same physical origin,
      interference of the incoming and outgoing waves. Consider the
      Schwarzschild case for simplicity. It is convenient to rewrite
      the radial wave function (\ref{R}) in a more formal abstract
      notation
\begin{eqnarray}
  \label{inout}
|  \, \phi \,\rangle = |\,\mathrm{in} \,\rangle + \mathcal{R} \, 
|\, \mathrm{out}\, \rangle~.
      \end{eqnarray}
      This notation assumes that the time-dependent factor is included
      in the wave function, i. e. $|\, \phi \, \rangle = \exp (-i
      \varepsilon t) \phi(r)$, where $\phi(r)$ is given in (\ref{R}).
      The two terms on the right-hand side of (\ref{R}) give the
      corresponding terms in (\ref{inout}) that can be conveniently
      written using Kruskal coordinates (\ref{U}),(\ref{V}) as
      \begin{eqnarray}
        \label{i}
 |\, \mathrm{in} \, \rangle ~ & =& \exp[-i \varepsilon\, \ln(V^2)\,]~,
\\         \label{o}
 |\, \mathrm{out} \, \rangle& =& \exp[~~i \varepsilon\, \ln(U^2)\,]~.
      \end{eqnarray}
      We restrict our discussion to the events that take place in the
      vicinity of the horizon where the semiclassical description
      holds, justifying (\ref{i}),(\ref{o}). The classical trajectory
      that corresponds to the incoming wave $|\,\mathrm{in}\,\rangle$
      follows from the equation of motion $\partial_\varepsilon S =0$,
      where the action reads $S = \varepsilon\, \ln(V^2)$.
      Therefore the ingoing trajectory is described by equation
      $V=const$.  Similarly the outgoing wave $|\, \mathrm{out}\,
      \rangle$ in the vicinity of the horizon corresponds to the
      classical trajectory $U=const$.  In $r,t$ variables these two
      trajectories are presented in (\ref{r-1}) for the outside
      region.
      
\begin{figure}[b]
\centering
\includegraphics[height=8cm,keepaspectratio=true]{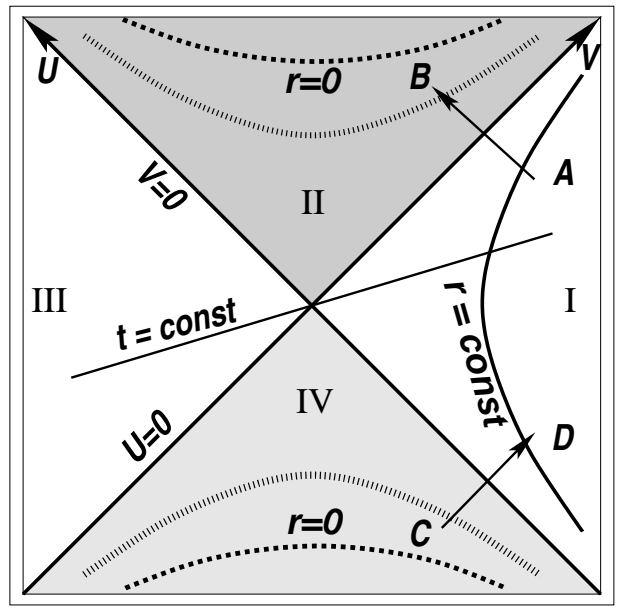}
\caption{\it 
  Kruskal coordinates. Areas I and III represent two identical copies
  of the outside region; II, IV show two inside regions. Hyperbolic
  curves $UV = const$ describe condition $r=const$, the dotted curve
  shows location of $r=0$, the inclined straight line presents
  condition $t=const$. The direction of time flow in I and III is
  opposite. The incoming particle follows {\it AB} crossing the
  horizon $U=0$ and residing in II. The outgoing particle {\it CD}
  escapes from IV crossing the horizon $V=0$ and coming to I.  Areas
  II and IV are not connected, which ensures classical confinement in
  II.  The wave function (\ref{R}) or (\ref{inout}) describe mixing of
  events that correspond to incoming and outgoing classical
  trajectories ({\it AB} and {\it CD}), resulting in phenomena of
  reflection and radiation.  }
\label{one}
\end{figure}

      Fig. \ref{one} shows classical trajectories in Kruskal
      coordinates.  This graphical presentation emphasizes the
      unexpected, nontrivial nature of the interference between the
      incoming and outgoing waves in (\ref{inout}). A particle that
      follows the incoming trajectory has no classically allowed
      chance to switch to the outgoing trajectory in the classical
      approximation.  Fig. \ref{one} visualizes this argument, showing
      that inside the event horizon the incoming and outgoing
      trajectories belong to different regions of the $U-V$ plane.
      Thus the incoming and outgoing trajectories seem to be
      completely unrelated. However, equation (\ref{inout}) indicates
      that on the quantum level there arises a connection between the
      incoming and outgoing waves. It manifests itself as the
      interference of these waves in the wave function.
      \noindent 
      We verified this statement above for the outside region $r>1$,
      but it holds for the inside region as well.  Indeed, Kruskal
      coordinates in Eqs.(\ref{i}),(\ref{o}) show that the logarithmic
      singularity of the wave function does not depend on the sign of
      $U$ and $V$, i.e. it exists on both sides of the horizon.
      Therefore inside the horizon one can use same method that we
      used above for the outside region, which leads to the same
      result that remains valid on both sides of the horizon: the
      incoming and outgoing waves do interfere in the wave function
      (\ref{inout}), see also discussion in Ref.{\cite{kuchiev_03}.

      We discussed in Sections \ref{reflection} and
      \ref{reflectionGen} the physical manifestation of this
      interference for the outside region, claiming that it leads to
      the reflection of the probing incoming particle from the event
      horizon. Let us now consider the physical manifestation of this
      interference for the region inside the horizon. The classical
      ingoing trajectory $V=const$ describes here the motion towards
      the black hole center, the outgoing $U=const$ trajectory
      describes the motion that eventually brings the particle from
      the inside region, over the horizon, into the outside region
      $r>1$. If a particle follows the ingoing classical trajectory
      then, as mentioned above, there is no classical way for it to
      switch to the outgoing trajectory and escape into the outside
      region. However, (\ref{inout}) shows that the perception based
      on the classical picture is not completely correct. In the
      quantum wave function the proper ingoing wave
      $|\,\mathrm{in}\,\rangle$ is mixed with the proper outgoing wave
      $|\,\mathrm{out}\,\rangle$. This mixing indicates that the
      particle that moves towards the black hole center in the inside
      region has a finite chance to simultaneously populate the
      outgoing wave that brings it to the outside region. Thus there
      is a finite probability for the particle to escape from the
      region inside the horizon into the outside region.
      
      Let us calculate this probability. Suppose that there is a
      particle confined in the inside region.  Assume that this
      particle occupies a state with the quantum numbers
      $\varepsilon,L,M$ moving from the horizon deeper inside the
      black hole, eventually aiming at the singularity at the origin
      \footnote{Generally speaking, to be certain that the particle is
        located in the inside region one needs to describe its motion
        with the help of the wave packet that propagates from the
        horizon into the deeper region. However, for our purposes it
        suffices to take into account only one wave with the given
        quantum numbers $\varepsilon,L,M$. Proving below that each
        wave of the packet has a chance to escape from the inside
        region, we prove simultaneously that the wave packet can
        escape as well.}.  According to the above discussion one
      should describe this particle by the wave function
      (\ref{inout}), which shows that there is an admixture of the
      outgoing wave. The probability to populate this wave is
      $\mathcal{P} =|\mathcal{R}|^2$.  Following the classical
      outgoing trajectory, which corresponds to this wave, the
      particle can reach the event horizon and therefore can escape
      into the outside world. 
      
      Thus there exists the probability that the particle escapes
      $\mathcal{P}_\mathrm{esc} \propto |\mathcal{R}|^2=\mathcal{P}$.
      We can be more specific.  We know that the wave that reaches the
      event horizon is partially reflected. According to (\ref{cro})
      the probability of reflection equals $\mathcal{P}_\mathrm{cr}=
      1-\mathcal{P}$.  We proved this result when we considered the
      scattering process that takes place in the outside region. One
      can verify that this result holds when we consider the
      scattering that takes place for the wave that comes to the
      horizon on its way from inside-out as well.  Combining the two
      factors, the probability to populate the outgoing wave, and the
      probability to cross the event horizon we conclude that the
      probability for the particle to escape into the outside world
      equals $\mathcal{P}_\mathrm{esc}= \mathcal{P} ( 1 - \mathcal{P}
      )$.  It is instructive to compare this result with the
      probability of the particle to be absorbed.  Suppose we have an
      incoming particle in the outside region in a state described by
      the wave function (\ref{inout}) with the same quantum numbers
      $\varepsilon,L,M$. The probability for this particle to populate
      the ingoing wave in (\ref{inout}) is unity, therefore the
      probability to be absorbed $\mathcal{P}_\mathrm{abs}$ into the
      inside region equals the probability to cross the event horizon
      (\ref{cro}), which gives $\mathcal{P}_\mathrm{abs} = 1 -
      \mathcal{P}$. We can consider now the ratio of the probability
      for a particle to escape from the inside region to the
      probability to be absorbed
      \begin{eqnarray}
        \label{escabs}
        \frac{ \mathcal{P}_\mathrm{esc} }
{ \mathcal{P}_\mathrm{abs} } =\mathcal{P} =\exp\left(
-\frac{\varepsilon}{kT} \right)~.
      \end{eqnarray}
      Discussing the probabilities above we considered only those
      factors that originate directly from the wave function
      (\ref{inout}). The physical probabilities include also
      additional normalization factors related to the flux of
      particles and the surface area of the event horizon. However,
      these additional factors are canceled out in the ratio
      (\ref{escabs}), which presents therefore the result for the
      ratio of the two physical rates, emittance and absorption. It
      states that the ratio of the emittance and absorption rates
      coincides with the conventional temperature factor that
      describes the ratio of these rates for the black body with the
      temperature $T$.  This means that if the black hole is put
      inside the thermostat with the temperature $T$, then it
      remains in equilibrium with it.  One concludes therefore that
      (\ref{escabs}) indicates that the black hole possesses the
      temperature $T$ radiating as a black body with this temperature,
      as was first discovered by Hawking
      \cite{Hawking_1974,Hawking_1975} using different arguments.
      
      There is a conventional physical explanation for the Hawking
      process that refers to the creation of pairs.  The gravitational
      field in the vicinity of the horizon creates a pair, then a
      particle goes into the outside world, while its anti-partner is
      absorbed by the black hole. This explanation of the process
      needs endeavor to approve the fact that the antiparticle brings
      into the black hole the negative amount of energy that
      compensates the energy of the created particle.
      Eq.(\ref{inout}) suggests an alternative simple explanation. The
      radiation happens because the particle confined inside the horizon
      can escape into the outside world. This point of view has
      automatically accounts for the reduction of the mass of the
      black hole; when the particle escapes from the black hole it
      contributes to the mass of the black hole no more. 

      Summarizing, we verified that both the reflection and the
      Hawking radiation stem from the interference of the incoming and
      outgoing waves in the wave function (\ref{inout}).

\section{Discussion and conclusion}
     \label{discussion}
     
     The existence of the event horizon that separates the outside and
     inside regions is the main property of black holes. It is well
     known that one can always choose the coordinate frame that makes
     the metric smooth on the horizon.  Correspondingly, the classical
     equations of motion for a probing particle in these coordinates
     are also smooth on the horizon.  From this fact follows a known
     conclusion: a probing particle that follows the classical
     trajectory on its way to the black hole crosses the horizon
     quite smoothly, but after that will be forced to stay inside
     forever.  However, quantum corrections influence the fate of this
     particle.  Presented arguments indicate that the horizon
     makes a strong impact on the wave function of a probing particle.
     It manifests itself in the form of interference, mixing of the
     incoming and outgoing waves in the wave function (\ref{inout}).
     Without this mixing the incoming wave crosses the event horizon
     quite uneventfully, in accord with similar smooth transition
     through the horizon of the classical trajectory. The mentioned
     mixing indicates that the incoming wave inevitably incorporates
     some admixture of the outgoing wave with interesting
     consequences, that are discussed below.  But first let us recall
     some details related to the interference {\it per se}. We
     verified the existence of the mixing, showed that it happens due
     to events localized on the horizon and calculated its magnitude
     using the semiclassical approximation.  The central role played
     the classical action for a probing particle that possesses the
     logarithmic singularity on the horizon.
     
     Importantly, this singularity persists in any coordinate frame,
     it exists even in those coordinates in which the metric is smooth
     on the horizon, for example, in Kruskal coordinates for the
     Schwarzschild metric. Using the analytical continuation of the
     semiclassical wave function in the vicinity of the horizon we
     found the coefficient $\mathcal{R}$ that describes the mixing
     of the incoming and outgoing waves in the wave function
     (\ref{inout}).  This coefficient possesses a typically
     semiclassical nature for a classically forbidden quantity,
     \begin{eqnarray}
       \label{A}
       |\mathcal{R}| =\exp\left(-\frac{\mathcal{A}}{\hbar} \right)~,
     \end{eqnarray}
     where $A$ has the meaning of some effective classical action. For
     example, for the Schwarzschild geometry of the black hole
     $\mathcal{A}=\varepsilon \,\tau$, where $\tau$ has the dimension
     of time with the typical value $\tau=2 \pi r_g/c$. In the
     classical limit $\hbar \rightarrow 0$ the mixing (\ref{A})
     disappears.  Thus, from the point of view of the classical
     approximation the physical manifestations of quantum interference
     look unusual.  Having said that, it is necessary to point out
     that in more common scattering situations there is nothing
     unusual about the interference between the incoming and outgoing
     waves, on the contrary, it is quite normal. The point is that
     black holes are very special. They are supposed to absorb very
     well everything incoming, therefore naively there should exist
     only the incoming wave that describes the particle that
     approaches the horizon.  From this perspective the existence of
     the interference and, consequently, existence of the reflected
     outgoing wave is surprising.
     
     There were discussed two effects that originate from the
     interference between the incoming and outgoing waves. One of them
     is a novel effect, reflection. For any particle that approaches
     the event horizon there is a finite probability to bounce back,
     into the outside world. The probability of reflection depends on
     the energy $\varepsilon$ of the incoming particle and the
     temperature $T$ of the black hole. For $\varepsilon < T$ the
     black hole behaves as a reflector, which is unusual.
 
     Another effect that follows from the interference of the incoming
     and outgoing waves is the well known phenomenon of the Hawking
     radiation. The suggested new explanation for this effect is
     simple and appealing. The radiation happens because when the
     incoming particle is confined in the inside region, it still
     maintains an opportunity to escape back into the outside world.
     This fact changes the perception of the event horizon.
     Conventional arguments claim that when the incoming particle
     comes into the inside region, it stays there forever, the horizon
     is impassable for the backward transition. This argument,
     however, holds only in the classical approximation. Quantum
     corrections make the horizon partially transparent, the particles
     can cross it and go away creating the Hawking radiation spectrum
     of the black hole.
     
     Both the radiative and reflective abilities of black holes arise
     from quantum corrections, both these processes are governed by
     the Hawking temperature of the black hole, but experimentally
     they are well distinguishable. The reflected flux depends on the
     nature, flux and spectrum of incoming particles, as well as on
     the black hole properties, while the radiation is governed
     entirely by the black hole. The radiation phenomenon provides
     support for important thermodynamics properties of black holes.
     The suggested new approach to the origins of the radiation may
     help to look anew at the thermodynamics properties of black holes
     as well, but this topic lies ahead.
     
     In conclusion, black holes are capable of reflection, this effect
     has a common physical origin with the Hawking radiation.
    
Discussions with V.V.Flambaum are appreciated. This work was supported
by the Australian Research Council.

%\bibliography

  \end{document}